# Glass phase and other multiple liquid-to-liquid transitions resulting from two-liquid phase competition

Robert F. Tournier[1,2]

[1]Univ. Grenoble Alpes, Inst. NEEL, F-38042 Grenoble Cedex 9, France

[2]CNRS, Inst. NEEL, F-38042 Grenoble Cedex 9, France
E-mail address: robert.tournier@neel.cnrs.fr

**Abstract**: Melt supercooling leads to glass formation. Liquid-to-liquid phase transitions are observed depending on thermal paths. Viscosity, density and surface tension thermal dependences measured at heating and subsequent cooling show hysteresis below a branching temperature and result from the competition of two-liquid phases separated by an enthalpy difference depending on temperature. The nucleation classical equation of these phases is completed by this enthalpy saving existing at all temperatures. The glass phase thermodynamic parameters and their thermal variation have already been determined in such a two-liquid model. They are used at high temperatures to predict liquid-to-liquid transitions in some metallic glass-forming melts.

## 1-Introduction

Phenomena of branching thermal properties of melts measured at heating and subsequent cooling have been observed for a long time and explained by the irreversible breakdown of a metastable micro-heterogeneous state of melts inherited from the original crystal sample or appearing in the process of melting [1-3]. This breakdown happens when heating up to the homogenizing temperature of the melts, which is often close to the branching temperature. The recent observations of liquid-to-liquid phase transitions (LLPT) in well-known metallic glass-forming melts raise the question of the competition between two homogeneous liquid phases numbered 1 and 2 separated by an enthalpy difference depending on the temperature [4-10]. The glass phase formation is often viewed as being due to a true thermodynamic transition. Various microscopic models and experiments prove its existence at $T_g$ [11-22]. The nucleation classical equation has been completed by introducing enthalpy savings $\varepsilon_{1s} \times \Delta H_m$ and $\varepsilon_{gs} \times \Delta H_m$, respectively associated with growth nucleus formation giving rise to crystallization in Phase 1 above $T_g$ and Phase 2 below $T_g$, where $\Delta H_m$ is the melting heat per g-atom [22]. The enthalpy saving $\Delta \varepsilon_{1g} \times \Delta H_m$ associated with the glass formation is then equal to $(\varepsilon_{1s} - \varepsilon_{gs}) \times \Delta H_m$. The energy saving coefficients $\varepsilon_{1s}$ and $\varepsilon_{gs}$ are linear functions of $\theta^2 = (T-T_m)^2/T_m^2$, as shown by a study of supercooling rate maxima of liquid elements [23] and are equal to $\varepsilon_{1s0}$ and $\varepsilon_{gs0}$ at $T_m$. The minimum value 0.217 of $\varepsilon_{1s0}$ and $\varepsilon_{gs0}$ determined in many liquid elements at their melting temperature $T_m$ corresponds to the Lindemann coefficient 0.103 [24]. The first-order transition to an ultrastable glass of confined liquid helium under pressure has been successfully described using $\varepsilon_{1s0} = \varepsilon_{gs0} = 0.217$ [25]. The transformation temperature of glasses in ultrastable phases having higher density has been defined as a function of an excess of frozen enthalpy $\Delta \varepsilon$. The enthalpy of the ultrastable phase attains the minimum and the density the maximum when $\Delta \varepsilon$ is equal to $\Delta \varepsilon_{1g0} = \varepsilon_{1s0} - \varepsilon_{gs0}$ at $T_m$. The time dependence of $T_g$ is explained by a weak excess enthalpy increasing $T_g$ and relaxing after cooling [26].

In this paper, a positive sign of $\Delta\varepsilon_{lg} = \varepsilon_{ls}-\varepsilon_{gs}$ shows that Phase 1 is favored and a negative value that it is Phase 2. The glass phase below $T_g$ corresponds to a liquid in Phase 2. The transformation of Phase 2 in a denser phase is not included in the enthalpy changes in order to simplify the presentation. The coefficients $\varepsilon_{ls}$ and $\varepsilon_{gs}$, being functions of $\theta^2$, are used above $T_m$. The homogeneous nucleation temperatures $T_{n-}$ below $T_m$ and $T_{n+}$ above $T_m$ of Phases 1 and 2 are solutions of the classical nucleation equation completed by the enthalpy saving $\Delta\varepsilon_{lg}\times\Delta H_m$ and lead to first-order LLPT accompanied by latent heats equal to $\Delta\varepsilon_{lg}\times\Delta H_m$. Two branching temperatures $T_{Br}$ of thermal properties are defined for $\Delta\varepsilon_{lg}=0$. The overheating and quenching temperatures $T_{q1}$ and $T_{q2}$ leading to the freezing of enthalpy excesses equal to $\Delta\varepsilon_{lg0}=\varepsilon_{ls0}-\varepsilon_{gs0}$ and $2\times\Delta\varepsilon_{lg0}$ are determined. The nucleation temperature of Phase 1 in the presence of an enthalpy excess $\Delta\varepsilon_{lg0}$ is also calculated and called $T^{ex}_{n-}$. The theoretical solution is attained with a superheating up to $T_{ts}$, where $\varepsilon_{gs}$ becomes equal to zero. The enthalpy saving coefficients $\varepsilon_{ls}$ and $\varepsilon_{gs}$ are counted from this t solution. Some of these calculated temperatures are successfully compared with various experimental determinations. The existence of a weak enthalpy excess relaxing at $T_g$ in Phase 2 inducing a time dependence of the glass transition is demonstrated.

## 2-Basic equations

The nucleation equation is given by (1):

$$\Delta G = \frac{4\pi R^3}{3}\Delta H_m/V_m \times (\theta-\varepsilon) + 4\pi R^2(1+\varepsilon)\sigma_1\Delta H_m/V_m \tag{1}$$

where $\Delta G$ is the Gibbs free energy change associated with the formation of a spherical growth nucleus of radius R, $\varepsilon$ being a fraction of the melting enthalpy $\Delta H_m$, $V_m$ the molar volume and $\theta = (T-T_m)/T_m$ the reduced temperature. The critical nucleus can give rise to crystallization or to Phase 1 or Phase 2 according to the value of the coefficient $\varepsilon$. The new surface energy is $(1+\varepsilon)\times\sigma_1$ instead of $\sigma_1$. The classical equation is obtained for $\varepsilon=0$ [27].

The homogeneous nucleation temperatures are $\theta_{n-}=(\varepsilon-2)/3$ for $\theta<0$ and $\theta_{n+}=\varepsilon$ for $\theta>0$. The nucleation temperature $\theta_{n+}=\varepsilon$ for $\theta>0$ has not been used up to now. It can be found in [22 above (13)]. The coefficients $\varepsilon_{ls}$ and $\varepsilon_{gs}$ are values of $\varepsilon(\theta)$ leading to a nucleus formation having the critical radius for crystal growth in Phase 1 and Phase 2 respectively.

$$\varepsilon_{ls}(\theta) = \varepsilon_{ls0}(1-\theta^2\times\theta_{0m}^{-2}), \tag{2}$$

$$\varepsilon_{gs}(\theta) = \varepsilon_{gs0}(1-\theta^2\times\theta_{0g}^{-2})+\Delta\varepsilon, \tag{3}$$

where $\Delta\varepsilon$ is a coefficient of enthalpy excess which may be frozen after quenching the melt [26]. Equations (2) and (3) are respected at the homogeneous nucleation temperatures of Phase 1 and Phase 2 and (4) is

deduced to determine the nucleation temperature $\theta_{n-}$ of the glass phase called Phase 2:

$$\theta_{n-}^2 \times \varepsilon_{gs0} \times \theta_{0g}^{-2} + 3\theta_{n-} + 2 - \varepsilon_{gs0} - \Delta\varepsilon = 0 \tag{4}$$

The solutions for $\theta_{n-}$ are given by (5):

$$\theta_{n-} = (-3 \pm \left[9 - 4(2 - \varepsilon_{gs0} - \Delta\varepsilon)\varepsilon_{gs0} / \theta_{0g}^2\right]^{1/2})\theta_{0g}^2 / (2\varepsilon_{gs0}) \tag{5}$$

The highest value of $\theta_{n-}$ is chosen with sign + in (5). The glass transition reduced temperature $\theta_{n-} = \theta_g$ is obtained for $\Delta\varepsilon = 0$. A second $\theta_{n-}$ is equal to $a \times \theta_g/1.5$ for a value of $\Delta\varepsilon$ leading to the nucleation temperature of Phase 1 in Phase 2 and to the homogeneous nucleation temperature of crystallization in Phase 1. Another value of $\theta_{n-}$ equal to $\theta^{ex}_{n-}$ is obtained for $\Delta\varepsilon = \varepsilon_{ls0} - \varepsilon_{gs0}$ as discussed below.

An enthalpy excess $\Delta\varepsilon \times \Delta H_m$ depending on the quenching temperature can be frozen in Phase 2 by rapid cooling. Two reduced temperatures $\theta_{q1}$ and $\theta_{q2}$ of quenching leading to $\Delta\varepsilon = (\varepsilon_{ls0} - \varepsilon_{gs0}) = \Delta\varepsilon_{lg0}$ and to $\Delta\varepsilon = 2 \times \Delta\varepsilon_{lg0}$ are calculated. They are defined for $\Delta\varepsilon_{lg} = 0$ in (10). An overheating of Phase 1 above $\theta_{q1}$ and a subsequent slow cooling down to $\theta_{q1}$ would have to produce a new LLPT leading to Phase 2 at this temperature. The enthalpy excess coefficient $\Delta\varepsilon = \Delta\varepsilon_{lg0}$ is privileged because it gives rise to the most ultrastable glass phase at a vapor deposition temperature $T_{sg}$ having the highest enthalpy saving [26]. All other values of $\Delta\varepsilon$ smaller than $2 \times \Delta\varepsilon_{lg0}$ lead to deposited ultrastable phase having larger enthalpy. It is inferred that the enthalpy excess $\Delta\varepsilon_{lg0} \times \Delta H_m$ maximizing the enthalpy saving of the ultrastable glass phase is frozen in glasses that are superheated in the temperature window delimited by $\theta_{q1}$ and $\theta_{q2}$ and rapidly cooled at low temperatures.

In fragile liquids, $\theta_{0m}$ being larger than -2/3 [22], the coefficients $\varepsilon_{ls0}$ and $\theta_{0m}$ in Phase 1 are given by (6) and (7), with "a" being unknown and $\theta_{n-}$ being the homogeneous nucleation reduced temperature of Phase 1 in Phase 2 and of growth nuclei giving rise to crystallization in Phase 1 after a very long time of relaxation:

$$\varepsilon_{ls0} = \varepsilon_{ls}(\theta = 0) = 1.5 \times \theta_{n-} + 2 = a \times \theta_g + 2, \tag{6}$$

where $\theta_{n-}$ is equal to $a \times \theta_g/1.5$. The coefficient $\varepsilon_{ls0}$ is maximized by (6) [28-30].

$$\theta_{0m}^2 = \frac{8}{9}\varepsilon_{ls0} - \frac{4}{9}\varepsilon_{ls0}^2. \tag{7}$$

The coefficients $\varepsilon_{gs0}$ and $\theta_{0g}$ are given at $T_g$ and below $T_g$ by (8) and (9):

$$\varepsilon_{gs0} = \varepsilon_{gs}(\theta = 0) = 1.5 \times \theta_g + 2, \tag{8}$$

$$\theta_{0g}^2 = \frac{8}{9}\varepsilon_{gs0} - \frac{4}{9}\varepsilon_{gs0}^2. \tag{9}$$

The reduced temperature $\theta_g$ is the homogeneous nucleation reduced temperature of Phase 2 in Phase 1 and of growth nuclei giving rise to crystallization in Phase 2 after a very long time of relaxation. The coefficient $\varepsilon_{gs0}$ is also maximized by (8) [28-30].

The enthalpy saving coefficient $\Delta\varepsilon_{1g}$ of Phase 1 with regard to Phase 2 is given by (10) for $\Delta\varepsilon = 0$:

$$\Delta\varepsilon_{1g}(\theta) = (\varepsilon_{ls} - \varepsilon_{gs}) = \varepsilon_{ls0} - \varepsilon_{gs0} + \Delta\varepsilon - \left(\theta^2\varepsilon_{ls0}/\theta_{0m}^2 - \theta^2\varepsilon_{gs0}/\theta_{0g}^2\right) \tag{10}$$

Phase 1 is favored when $\Delta\varepsilon_{1g} > 0$ and Phase 2 when $\Delta\varepsilon_{1g} < 0$ for $\Delta\varepsilon = 0$. The phase changes occur at the homogeneous nucleation temperatures.

The reduced temperatures $\theta_{Br}$ below which thermal properties are irreversible are determined for $\Delta\varepsilon_{1g} = 0$ in (11):

$$\theta_{Br} = \pm\left[(\varepsilon_{ls0} - \varepsilon_{gs0})/(\varepsilon_{ls0}\theta_{0m}^{-2} - \varepsilon_{gs0}\theta_{0g}^{-2})\right]^{1/2} \tag{11}$$

The specific heat change $\Delta C_p(T)$ from Phase 2 to Phase 1 is given in (12) by $d\Delta\varepsilon_{1g}/dT$ multiplied by $\Delta H_m$ [22].

$$\Delta C_p(T) = 2\frac{T - T_m}{T_g - T_m}\frac{\Delta H_m}{T_m}\left[\frac{9}{4a} - \frac{9}{6}\right] \tag{12}$$

The parameter "a" in (6) is determined by the specific heat jump at $T_g$ when $T_m$ and $\Delta H_m$ are known. A lot of glasses follow the scaling law (6) in which $a = 1$, leading to $\Delta C_p(T_g) = 1.5 \times \Delta H_m/T_m$ [22].

The homogeneous nucleation reduced temperature of Phase 2 in Phase 1 above $T_m$ is given by (13) respecting $\Delta\varepsilon_{1g} = \theta_{n+}$:

$$\left(\frac{\varepsilon_{ls0}}{\theta_{0m}^2} - \frac{\varepsilon_{gs0}}{\theta_{0g}^2}\right)\theta_{n+}^2 + \theta_{n+} - (\varepsilon_{ls0} - \varepsilon_{gs0} + \Delta\varepsilon) = 0 \tag{13}$$

The solution for $\theta_{n+}$ is given by (14) for $\Delta\varepsilon = 0$:

$$\theta_{n+} = \frac{-1 + (1 + 4(\varepsilon_{ls0} - \varepsilon_{gs0})(\varepsilon_{ls0}\theta_{0m}^{-2} - \varepsilon_{gs0}\theta_{0g}^{-2}))^{1/2}}{2(\varepsilon_{ls0}\theta_{0m}^{-2} - \varepsilon_{gs0}\theta_{0g}^{-2})} \tag{14}$$

Phase 1 disappears when $\varepsilon_{ls}$ in (2) becomes equal to zero. The reduced temperature $\theta_{ts}$ above which all residual Phase 2 clusters cannot survive in a unique liquid and a theoretical solution of all atoms is given by (15) corresponding to $\varepsilon_{gs} = 0$ in (3):

$$\theta_{ts} = \theta_{0g} \tag{15}$$

## 3-Application to glass-forming melts obeying scaling laws

A broad fraction of metallic and non-metallic glasses have a specific heat jump $\Delta C_p(T_g) = 1.5 \times \Delta S_m$, $\Delta S_m = \Delta H_m/T_m$ being the melting entropy. This value corresponds to a =1 in [6,12]. The Vogel–Fulcher–Tammann temperature of many polymers follows a scaling law [31] in agreement with (6). The quantity $\Delta \varepsilon_{lg0}$ is equal $-0.5 \times \theta_g$. The following LLPT are expected in this family:

| | | |
|---|---|---|
| $\theta_{n-} = 2/3 \times \theta_g$, | $\varepsilon_{ls} - \varepsilon_{lg} = -0.16666 \times \theta_g$, | (16) |
| $\theta^{ex}_{n-} = 0.42222 \times \theta_g$ | $\varepsilon_{ls} - \varepsilon_{lg} = -0.3663 \times \theta_g$ | (17) |
| $\theta_{n+} = -0.38742 \times \theta_g$ | $\varepsilon_{ls} - \varepsilon_{lg} = -0.38742 \times \theta_g$ | (18) |
| $\theta_{Br} = \pm 0.8165 \times \theta_g$ | $\varepsilon_{ls} - \varepsilon_{lg} = 0$ | (19) |
| $\theta_{q1} = -1.1547 \times \theta_g$ | $\varepsilon_{ls} - \varepsilon_{lg} = \varepsilon_{ls0} - \varepsilon_{gs0} = -0.5 \times \theta_g$ | (20) |

The superheating temperature $\theta_{q2} = -\theta_g \times \sqrt{2}$ corresponds to $\Delta \varepsilon = 2 \times (\varepsilon_{ls0} - \varepsilon_{gs0}) = -\theta_g$.
Phase 1 disappears above a reduced temperature equal to $-\theta_{0m} = [(2+\theta_g) \times (-4 \times \theta_g/9)]^{1/2}$ in all glass-forming melts.
Phase 2 clusters disappear above $\theta_{ts} = [(1.5 \times \theta_g + 2) \times (-2 \times \theta_g/3)]^{1/2}$ in all glass-forming melts.

## 4-Application to some metallic glass-forming melts

### 4-1- $Ni_{77.5}B_{22.5}$

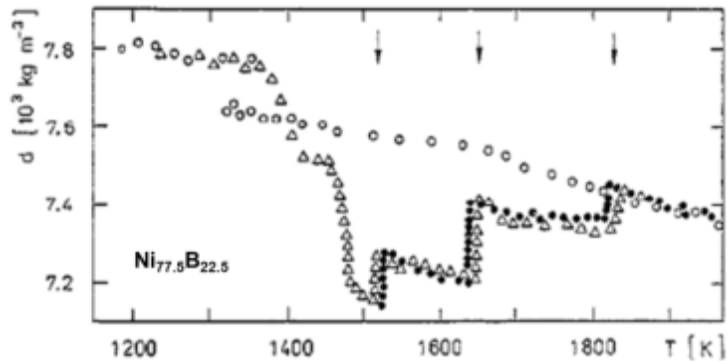

**Figure 1**: *Reprinted from [1]: Temperature dependence of the density d of Ni-22.5%B melt at heating after melting (●), subsequent cooling (0), and the second heating after crystallization of the sample and repeated melting (Δ). The arrows show the 'critical' temperatures at which the density instability is observed.*

Figure 1 shows the thermal variation of the density of Ni-22.5at%B melt. The melting temperature is $T_m$ =1361 K. The sample has been exposed to an annealing time of a few hours after melting. The density increases after such annealing above $T_m$. Stable density values indicated by arrows increase with increasing temperature. These steps are viewed as due to the appearance of microdomains enriched with one of the components in metallic glasses. A microheterogeneous state exists in liquid alloys and is revealed by prolonged relaxation time [1]. These new liquid states are initiated in Phase 1. This alloy undergoes a transition to a glass state after hyperquenching and is transformed from Phase 2 to Phase 1 after heating through $T_g$ and above $T_m$. The glass transition temperature $T_g$ would have to be equal to 617 K ($\theta_g$ =-0.547) in Table 1 as expected in strong glasses having $\varepsilon_{gs0}$=0.514 and $\varepsilon_{ls0}$ =1.099 [22]. The crystallization is then obtained by heating Phase 1. Various types of Phase 1 are formed by heating, leading to exothermic latent heats at 1520 and 1645 K after an endothermic jump occurring at 1450 K. The homogeneous nucleation of Phase 2 starts at $\theta_{n+}$ =0.348, $T_{n+}$ =1835 K with $\varepsilon_{ls0}$ ≅1.099 and an endothermic latent heat equal to 0.348×$\Delta H_m$. The density variation at $T_m$ is proportional to the latent heat of melting. The density step at $T_{n+}$=1835 K is, as expected, about 35% of the density jump at $T_m$. Phase 1 is, step by step, progressively homogenized by heating.

**Table 1 LLPT temperatures deduced from thermodynamic parameters and experiments**.
*All underlined values are both experimental and calculated values. 1- $T_m(K)$ the melting temperature of crystals, 2- $T_g(K)$ the glass transition temperature, 3- $\theta_g$ the reduced glass transition temperature, 4- $\theta_{Br}$ the branching reduced temperature below $T_m$ given in (11), 5- $T_{Br}(K)$ the branching temperature below $T_m$, 6- $\Delta\varepsilon_{lg}$ the difference ($\varepsilon_{ls} - \varepsilon_{gs}$) =0 at $T_{Br}$ below $T_m$, 7- $\varepsilon_{gs0}$ the enthalpy saving coefficient $\varepsilon_{gs}$ of Phase 2 at $T_m$, 8- $\theta_{0g}^2$ the square of the reduced temperature where $\varepsilon_{gs}$ is equal to zero, 9- $\Delta C_p(T_g)$ in $J.K^{-1}/g.atom$. the specific heat jump at $T_g$ defined in (12), 10- a the coefficient defined in (6), 11- $\varepsilon_{ls0}$ the enthalpy saving coefficient $\varepsilon_{ls}$ of Phase 1 at $T_m$, 12- $\theta_{0m}^2$ the square of the reduced temperature where $\varepsilon_{ls}$ is equal to zero, 13- $\Delta\varepsilon_{lg0}$ the difference between the enthalpy coefficients $\varepsilon_{ls0}$-$\varepsilon_{gso}$ at $T_m$, 14- $\theta_{n-}$ the reduced temperature of Phase 1 nucleation below $T_m$, 15- $T_{n-}(K)$ the nucleation temperature of Phase 1 below $T_m$, 16- $\Delta\varepsilon_{lg}(T_{n-})$ the difference in (10) between $\varepsilon_{ls}$ and $\varepsilon_{gs}$ at $T_{n-}$ for $\Delta\varepsilon$ =0, 17- $\theta^{ex}_{n-}$ the reduced nucleation temperature of Phase 1 below $T_m$ for $\Delta\varepsilon = \Delta\varepsilon_{lg0}$, 18- $T^{ex}_{n-}(K)$ the nucleation temperature of Phase 1 below $T_m$ for $\Delta\varepsilon =\Delta\varepsilon_{lg0}$, 19- $\Delta\varepsilon_{lg}(T^{ex}_{n-})$ the value of $\Delta\varepsilon_{lg}$ given in (10) for $\Delta\varepsilon =\Delta\varepsilon_{lg0}$, 20- $\theta_{n+}$ the reduced nucleation temperature of Phase 2 above $T_m$ given in (14), 21- $T_{n+}(K)$ the nucleation temperature of Phase 2 above $T_m$ given in (14), 22- $\Delta\varepsilon_{lg}(T_{n+})$ the difference $\Delta\varepsilon_{lg}$ at $T_{n+}$ given in (10) for $\Delta\varepsilon$ =0, 23- $\theta_{Br}$ the reduced branching temperature above $T_m$ given in (11), 24- $T_{Br}(K)$ the branching temperature above $T_m$, 25- $\Delta\varepsilon_{lg}(T_{Br})$ =0 the value of $\Delta\varepsilon_{lg}$ at $T_{Br}$ given by (10) for $\Delta\varepsilon$ =0, 26- $\theta_{q1}$ the reduced temperature of quenching leading to $\Delta\varepsilon = \Delta\varepsilon_{lg0}$ frozen in Phase 2 and of a new LLPT, 27- $T_{q1}(K)$ the temperature of Phase 2 quenching and of a new LLPT, 28- $\Delta\varepsilon_{lg}(T_{q1})$ the value of $\Delta\varepsilon_{lg}$ =-$\Delta\varepsilon_{lg0}$ at $T_{q1}$ for $\Delta\varepsilon$ =0 and equal to zero in (10) for $\Delta\varepsilon = \Delta\varepsilon_{lg0}$, 29- $\theta_{q2}$ the reduced temperature of quenching leading to $\Delta\varepsilon = \Delta\varepsilon_{lg0}$ frozen in Phase 2, 30- $T_{q2}(K)$ the temperature of quenching of Phase 2 where $\Delta\varepsilon_{lg}$ =-2×$\Delta\varepsilon_{lg0}$, 31- $\Delta\varepsilon_{lg}(T_{q2})$ the value of $\Delta\varepsilon_{lg}$=-2×$\Delta\varepsilon_{lg0}$ at $T_{q2}$ in (10) for $\Delta\varepsilon$ =0 and equal to zero for $\Delta\varepsilon =2\times\Delta\varepsilon_{lg0}$, 32- $T_{ts}(K)$ the temperature above which the melt becomes a true solution.*

|   |   | $Ni_{77.5}B_{22.5}$ | $(Fe..)_{96}Nb_4$ | Vit105 | Vit106 | Vit1 |
|---|---|---|---|---|---|---|
| 1 | $T_m$ | 1361 | 1410 | 1072 | 1115 | 965 |
| 2 | $T_g$ | 617 | 863 | 675 | 682 | 625 |
| 3 | $\theta_g$ | -0.547 | -0.388 | -0.370 | -0.388 | -0.352 |
| 4 | $\theta_{Br}$ | -0.547 | -0.317 | -0.302 | -0.304 | -0.254 |
| 5 | $T_{Br}$ | 617 | 963 | 748 | 776 | 720 |
| 6 | $\Delta\varepsilon_{lg}(T_{Br})$ | 0 | 0 | 0 | 0 | 0 |
| 7 | $\varepsilon_{gs0}$ | 0.514 | 1.418 | 1.444 | 1.417 | 1.472 |
| 8 | $\theta_{0g}^2$ | 1 | 0.367 | 0.357 | 0.367 | 0.346 |
| 9 | $\Delta C_p(T_g)$ |  | 10 |  | 16.0 | 23.6 |
| 10 | a |  | 1 | 1 | 0.918 | 0.780 |

| 11 | $\varepsilon_{ls0}$ | 1.099 | 1.612 | 1.63 | 1.644 | 1.725 |
|---|---|---|---|---|---|---|
| 12 | $\theta_{0m}^2$ | 0.444 | 0.278 | 0.268 | 0.260 | 0.211 |
| 13 | $\Delta\varepsilon_{lg0}$ | 0.585 | 0.194 | 0.185 | 0.227 | 0.254 |
| 14 | $\theta_{n-}$ | -0.666 | -0.259 | -0.247 | **-0.238** | **-0.183** |
| 15 | $T_{n-}$ | 454 | 1045 | 807 | **850** | **788** |
| 16 | $\Delta\varepsilon_{lg}(T_{n-})$ | -0.285 | 0.065 | 0.062 | 0.174 | 0.122 |
| 17 | $\theta^{ex}_{n-}$ | -0.318 | -0.164 | -0.157 | **-0.238** | -0.103 |
| 18 | $T^{ex}_{n-}$ | 929 | 1179 | 904 | **952** | 865 |
| 19 | $\Delta\varepsilon_{lg}(T^{ex}_{n-})$ | 0.388 | 0.142 | 0.157 | 0.174 | 0.212 |
| 20 | $\theta_{n+}$ | **0.348** | **0.150** | **0.143** | 0.162 | **0.157** |
| 21 | $T_{n+}$ | **1835** | **1622** | **1226** | 1296 | **1116** |
| 22 | $\Delta\varepsilon_{lg}(T_{n+})$ | **0.348** | 0.150 | 0.143 | 0.162 | **0.157** |
| 23 | $\theta_{Br}$ | 0.547 | 0.317 | 0.302 | **0.304** | **0.254** |
| 24 | $T_{Br}$ | 2105 | 1857 | 1396 | **1454** | **1210** |
| 25 | $\Delta\varepsilon_{lg}(T_{Br})$ | 0 | 0 | 0 | 0 | **0** |
| 26 | $\theta_{q1}$ | 0.749 | 0.448 | 0.428 | 0.430 | 0.359 |
| 27 | $T_{q1}$ | 2380 | 2042 | 1530 | 1594 | 1312 |
| 28 | $\Delta\varepsilon_{lg}(T_{q1})$ | $-\Delta\varepsilon_{lg0}$ | $-\Delta\varepsilon_{lg0}$ | $-\Delta\varepsilon_{lg0}$ | $-\Delta\varepsilon_{lg0}$ | $-\Delta\varepsilon_{lg0}$ |
| 29 | $\theta_{q2}$ | 0.907 | 0.549 | 0.524 | 0.526 | 0.440 |
| 30 | $T_{q2}$ | 2596 | 2184 | 1633 | 1702 | 1390 |
| 31 | $\Delta\varepsilon_{lg}(T_{q2})$ | $-2\Delta\varepsilon_{lg0}$ | $-2\Delta\varepsilon_{lg0}$ | $-2\Delta\varepsilon_{lg0}$ | $-2\Delta\varepsilon_{lg0}$ | $-2\Delta\varepsilon_{lg0}$ |
| 32 | $T_{ts}$ | 2722 | 2264 | 1712 | 1790 | 1532 |

A branching temperature is observed at $T_{n+}$ =1835 K [1] instead of the predicted value of 2105 K given in Table 1. The thermal properties of Phase 1 are reversible up to $T_{Br}$, where a new branching temperature is expected with a formation of Phase 2. Phase 2 clusters would disappear at $T_{ts}$ = 2722 K.

## 4.2- $(Fe_{71.2}B_{24}Y_{4.8})_{96}Nb_4$

Amorphous alloys of $(Fe_{71.2}B_{24}Y_{4.8})_{96}Nb_4$ have been prepared by fast cooling of the melt from the temperature range of 1573–1773 K. The melting temperature is 1410 K and the glass transition temperature 863 K [6]. All properties of this melt are described in Table 1 using (16–20) and assuming that scaling laws are followed and consequently a =1. All LLPT temperatures as well as the fraction of melting heat associated with these first-order transitions are predicted. An endothermic heat has been observed at 1615–1650 K (6) and corresponds to the nucleation temperature $T_{n+}$ =1622 K of Phase 2 in Phase 1 predicted in (18) with $\theta_g$ =-0.38794. The latent heat would have to be equal to 15% of the melting heat.

## 4.3- $Zr_{52.5}Cu_{17.9}Ni_{14.6}Al_{10}Ti_5$ (Vit 105)

The glass transition temperature is $T_g$ =675 K and the melting temperature $T_m$ =1072 K (T solidus). A very precise overheating threshold has been observed at 1225 K on the supercooling rate of Vit 105, as shown in Figure 2 [32]. There are two nucleation temperatures of the same crystalline phase in liquid Phase 1 and liquid Phase 2. The model predicts $T_{n+}$ =1225 K for the nucleation of Phase 2 with (18), $\theta_g$ =-0.37034 assuming a =1. All parameters given in Table 1 are calculated from (16–20). Using a specific heat jump measurement of $\Delta C_p (T_g) =17.1$ J/at.g (33), a =0.86, the predicted temperature $T_{n+}$ would be equal to 1243 K (+1.47 %).

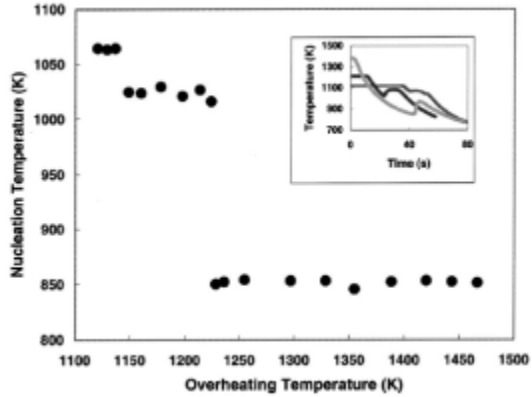

**Figure 2**: *Reprinted from [32], (Copyright 2004 American Institute of Physics). Nucleation temperature vs overheating temperature for $Zr_{52.5}Cu_{17.9}Ni_{14.6}Al_{10}Ti_5$ (Vit105). The cooling curves obtained from different levels of overheating are shown in the inset, in which the temperature at time t =0 s shows the level of overheating.*

### 4.4- $Zr_{57}Cu_{15.4}Ni_{12.6}Al_{10}Nb_5$ (Vit 106)

Two LLPT have already been obtained. The first one around 1450 K observes the specific volume as a function of temperature [7] below this branching temperature, the second one at 950–1000 K measures a structural crossover [8]. These two events correspond to $T^{ex}_{n-} =952$ K and $T_{Br} =1454$ K in Table 1. The calculation is made using a homogeneous nucleation temperature $T_{n1-}$ of crystallization in Phase 1 equal to 850 K, as observed by [8]. The corresponding values of a and $\Delta C_p (T_g)$ are 0.918 and 16 J/g.atom respectively. The structural crossover is calculated at 952 K because an overheating above 1600 K followed by rapid cooling has been used and results in freezing an enthalpy excess of $(\varepsilon_{ls0}-\varepsilon_{gs0}) \times \Delta H_m = 0.2265 \times \Delta H_m$.

### 4-5- $Zr_{41.2}Ti_{13.8}Cu_{12.5}Ni_{10}Be_{22.5}$ (Vit 1)

The glass transition of this melt is $T_g =625$ K and its melting temperature 965 K is chosen in the middle of the liquidus–solidus temperature window [34]. For the specific heat jump $\Delta C_p(T_g) =23.6$ J/g.atom of this fragile glass and $\Delta H_m =8200$ J/g.atom, the value of a is 0.78. The LLPT temperatures are deduced and given in Table 1. The thermal variation of $\Delta \varepsilon_{1g}$ given in (10) without frozen enthalpy ($\Delta \varepsilon =0$) is represented in Figure 3. Two branching temperatures are expected for $\Delta \varepsilon_{1g} =0$, the first one at 720 K and the second one at 1210 K. Phase 1 is the most stable phase between 720 and 1210 K. Phase 2 is the most stable phase below 720 K and above 1210 K.

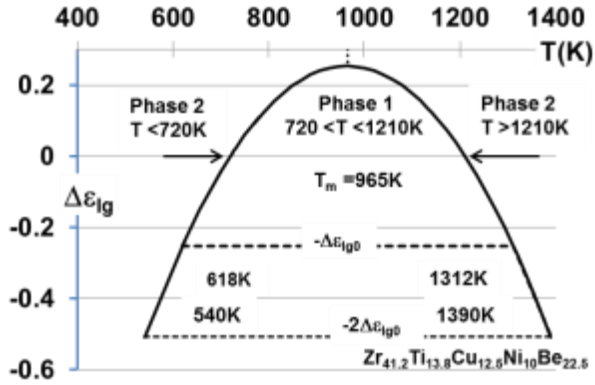

**Figure 3:** *Thermal variation of $\Delta\varepsilon_{lg}$ given by (10) for $\Delta\varepsilon = 0$. The most stable phase between 720 and 1210 K is Phase 1 and below 720 K and above 1210 K Phase 2. At the temperature $T_{q1} = 1312$ K, Phase 1 being not stable has an enthalpy excess of $\Delta\varepsilon_{lg0} \times \Delta H_m = (\varepsilon_{ls0} - \varepsilon_{gs0}) \times \Delta H_m$ as compared with Phase 2. Cooling the sample in Phase 1 instead of Phase 2 from above $T_{q1}$ down to $T_{q1}$ would lead to a LLPT from Phase 1 to Phase 2 because $\Delta\varepsilon_{lg}$ becomes equal to zero including $\Delta\varepsilon = \Delta\varepsilon_{lg0}$ in (10). At the temperature $T_{q2} = 1390$ K, Phase 1 has an enthalpy excess of $2 \times (\varepsilon_{ls0} - \varepsilon_{gs0}) \times \Delta H_m$.*

The homogeneous nucleation temperatures of Phase 2 are $T_g = 625$ K and $T_{n+} = 1116$ K. The branching temperature $T_{Br}$ has been observed at 1225 K on the viscosity and predicted at 1210 K. Figure 4 represents the branching of this quantity [5]. The viscosity has Newtonian behavior when the phase is the most stable.

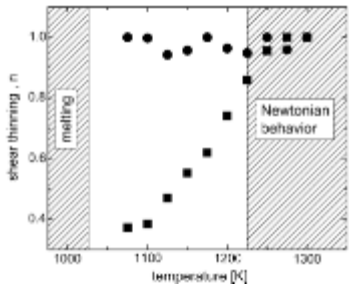

**Figure 4:** *Reprinted from [5]. Non-Newtonian behavior vs temperature of initially amorphous Vit1 for a first (□) and a second (○) temperature scan. After melting, the alloy exhibits a pronounced shear thinning dependence of the viscosity that decreases with increasing temperature. Above 1225 K, there is no longer a shear rate dependence of the viscosity. After cooling back to 1075 K and isothermal holding for 1 h, the material is still Newtonian over the investigated temperature range.*

The enthalpy coefficients $-\varepsilon_{ls}$ of Phase 1 and $-\varepsilon_{gs}$ of Phase 2 given by (2) and (3) are represented in Figure 5 between $T_g = 625$ K and 1225 K, including the latent heat coefficients equal to $(\varepsilon_{ls} - \varepsilon_{gs})$ at $T_{n-} = 788$ K and $T_{n+} = 1116$ K. Two LLPT have been observed at 760–830 K [35] and 1100–1200 K [4] as well as latent heats equal to about 900 J/g.atom [34] and 1100 J/g.atom respectively. The $(\varepsilon_{ls} - \varepsilon_{gs}) = 0.122 \times 8200 = 1000$ J/g.atom at 788 K and $0.157 \times 8200 = 1287$ J/g.atom are predicted in Table 1. The latent heat at 788 K is exothermic and corresponds to a transition from Phase 2 to Phase 1. That at 1116 K is endothermic and occurs at the nucleation temperature of Phase 2 in Phase 1. These two temperatures are not symmetrical as compared to $T_m$ =965 K. The latent heat at $T_{n+}$ is always larger than that at $T_{n-}$. Consequently, there is always a fraction of enthalpy which is not recovered below $T_{n-}$ in Phase 1.

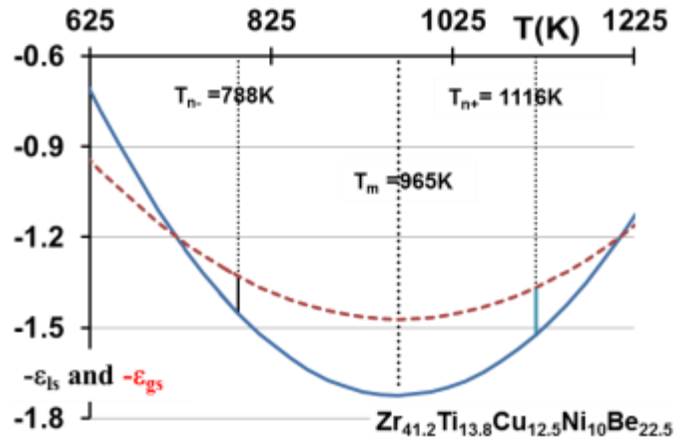

**Figure 5**: *The enthalpy coefficients $-\varepsilon_{ls}$ of Phase 1 (solid curve) and $-\varepsilon_{gs}$ of Phase 2 (dashed curve) vs temperature from 625 K up to 1225 K. Two LLPT are represented at 788 K and 1116 K. The first one is the homogeneous nucleation temperature of Phase 1 in Phase 2. The second one is that of Phase 2 in Phase 1.*

The enthalpy coefficients $-\varepsilon_{ls}$ of Phase 1 and $-\varepsilon_{gs}$ of Phase 2 given by (2) and (3) are represented in Figure 6 between $T_m$ = 965 K and $T_{ts}$ =1532 K, including latent heat coefficients equal to $(\varepsilon_{ls0}-\varepsilon_{gs0})$ at $T_{q1}$=1312 K and 2×$(\varepsilon_{ls0}-\varepsilon_{gs0})$ at $T_{q2}$=1390 K. A new LLPT is expected at $T_{q1}$ =1312 K after heating Phase 1 above this temperature and cooling it down to 1312 K. The temperature $T_{q2}$ =1390 K is indicated because it determines the upper limit of enthalpy excess leading to an ultrastable glass state after hyperquenching and liquid deposition on a substrate cooled below $T_g$ [26].

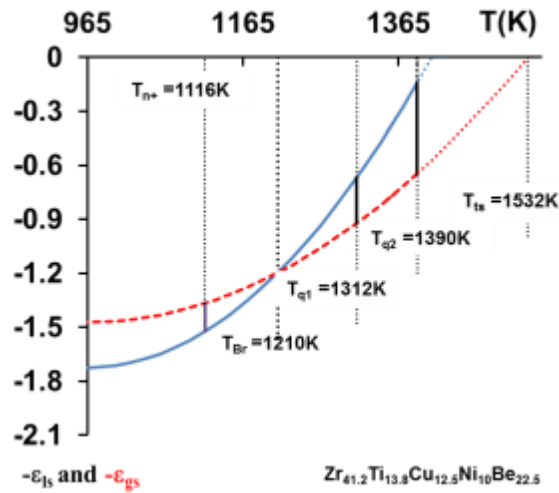

**Figure 6**: *Enthalpy coefficients $-\varepsilon_{ls}$ (solid curve) and $-\varepsilon_{gs}$ (dashed curve) of Phase 1 and Phase 2 given by (2) and (3) vs temperature from $T_m$ up to the true solution temperature $T_{ts}$ =1532 K. A new LLPT is expected at $T_{q1}$ =1312 K.*

## 5- Thermal paths leading to Phase 1 and Phase 2

Slow cooling below $T_m$ of Phase 1 and Phase 2 leads to only one crystal phase. New melting of crystals would have to stabilize Phase 1 above $T_m$ instead of Phase 2.

Quenching Phase 1 from a temperature less than $T_{n+}$ does not induce a LLPT at $T_{n-}$. A tendency to crystallization can exist at $T_{n-}$ because this temperature is also the homogeneous nucleation temperature of growth nuclei of crystals in Phase 1.

Heating Phase 1 up to $T_{n+}$ and above leads to Phase 2 through a LLPT [4,32]. A rapid cooling induces a LLPT at $T_{n-}$ which transforms Phase 2 into Phase 1 above $T_g$. Phase 1 is transformed into Phase 2 at $T_g$.

A rapid heating above $T_{q1}$ of Phase 2 followed by rapid cooling freezes an enthalpy excess in Phase 2 equal to $(\varepsilon_{ls0}-\varepsilon_{gs0})$ and produces a LLPT at $T^{ex}_{n-}$, the nucleation temperature of Phase 1 in Phase 2 [8]. A continuous cooling below $T^{ex}_{n-}$ of Phase 1 leads to $T_{n-}$ corresponding to the homogeneous nucleation temperature of some growth nuclei of crystals in Phase 1. Phase 1 is transformed into Phase 2 at $T_g$. With a new heating above $T_g$, Phase 2 is transformed into Phase 1 at $T_g$.

A rapid heating from the glass state up to $T_{n+}$ induces a LLPT at $T_{n+}$ transforming Phase 1 into Phase 2 [4].

## 6- Time dependence of the glass transition temperature $T_g$

The glass transition temperature $T_g$ depends on the heating and cooling rates, even if $T_g$ has a well-defined value at low rates [36]. Using (5), the glass transition temperature of Vit1 is plotted in Figure 7 as a function of the enthalpy excess coefficient $\Delta\varepsilon$ which has not been recovered in Phase 1. The $T_g$ can be increased from 625 to 720 K with $\Delta\varepsilon$ varying from zero to 0.0411. The temperature 720 K corresponds to the branching temperature close to $T_g$. The $T_g$ of Vit 1 increases up to 675 K by varying the heating rate from 0.0167 K/s to 6.66 K/s [34]. This increase corresponds to a value of $\Delta\varepsilon = 0.0115$. The enthalpy excess coefficient which has not been recovered at $T_{n-}$ is equal to $(0.157-0.122) = 0.035$ as shown in Table 1. Then, $T_g$ can still be increased with the heating rate. The time dependence of $T_g$ shows that the residual enthalpy excess is relaxed in Phase 2 below $T_g$ instead of Phase 1.

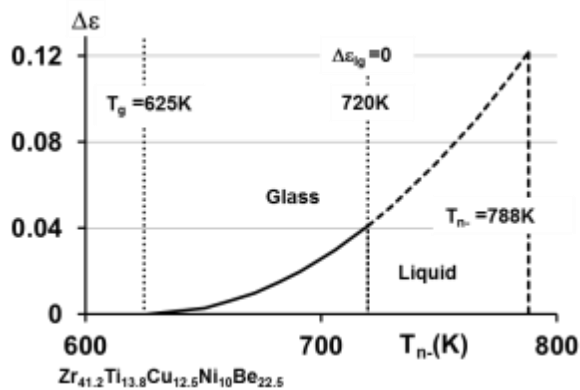

**Figure 7**: *The enthalpy excess coefficient $\Delta\varepsilon$ vs the glass transition up to 720 K as given in (5). Above 720 K, there are thermodynamic transitions associated with the presence of an enthalpy excess at $T_{n-} = 788$ K and $T^{ex}_{n-} = 865$ K. The other values of $\Delta\varepsilon$ above 720 K are not used. In all fragile glasses, $T_g$ is strongly dependent on small values of $\Delta\varepsilon$.*

## 7- Application to all liquids

This model can be applied to all liquids, even if they have never been transformed by quenching into glasses and if there is no proof of the existence of such LLPT in non-metallic glasses. The case of eutectic ($Al_{82.8}$-$Cu_{17.2}$) is examined [37]. The melting temperature $T_m$ is equal to 821.3 K and the formation temperature of Phase 2 is equal to 1603 K below $T_{0g} = 2\times T_m = 1642.6$ K ($\theta_{0g} = -1$ as in pure liquid elements) [24]. The value of $\varepsilon_{gs}$ in Phase 2 at T =1603 K ($\theta = 0.952$) is $0.217\times(1-\theta^2) = 0.02$. The highest condensation temperature of Phase 2 at 1603 K is accompanied by a branching of thermal properties because Phase 1 can be rapidly heated from $T_m$ up to 1603 K [37]. There is slope breaking at 1368 K corresponding to the melting of Phase 1 at $\theta = -2/3$ and the melt entrance into a solution without Phase 1. Phase 2 cluster formation occurs between 1603 and 1642.6 K. Phase 2 is induced at the homogeneous nucleation temperature determined by $\varepsilon_{gs} = 0.02$.

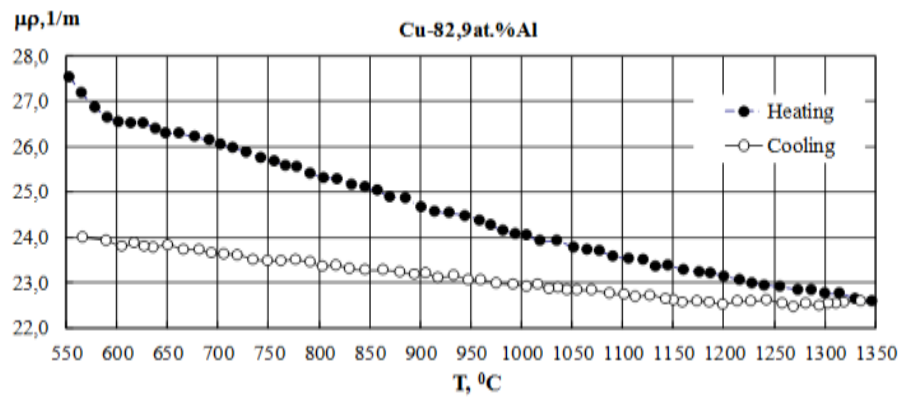

**Figure 8**: *Reprinted from [37] with permission of Springer. Temperature dependences of $\rho\mu$ ($\rho$ - density of the melt, $\mu$ - its gamma-quanta mass coefficient of absorption) for Al-Cu17.2 at% alloy.*

## Conclusions

Metallic glass and glass-forming melt properties are the result of competition between two liquid phases called Phase 1 and Phase 2 having an enthalpy difference varying with the square of the reduced temperature $\theta = (T-T_m)/T_m$. The transitions between these phases are governed by the classical equation of nucleation completed by this enthalpy saving. The liquid-to-liquid phase transitions (LLPT) of glass-forming melts are universal when they obey scaling laws and are determined knowing only $T_g$ and the melting temperature $T_m$ of crystallized phase. These glasses have a specific heat jump $\Delta C_p(T_g) = 1.5\times\Delta H_m/T_m$ from Phase 2 to Phase 1, heating the glass through $T_g$, and a homogeneous reduced nucleation temperature of Phase 1 equal to $\theta_g/1.5$ observed by cooling. The homogeneous nucleation temperature of Phase 1 in Phase 2 in other glass-forming melts with a specific heat jump $\Delta C_p(T_g)$ larger than $1.5\times\Delta S_m$ is equal to $a\times\theta_g/1.5$, where a is deduced from $\Delta C_p(T_g)$.

Four LLPT are predicted. Three have been observed up to now at the predicted temperatures in several glass-forming melts. Two nucleation temperatures of crystallized phases depending on supercooling of Phase 1 and Phase 2 were obtained in Vit105. The junction temperatures of Phase 1 and Phase 2 correspond to temperatures above which the thermal properties become reversible.

The LLPT of melts which have not been recognized, up to now, as glass-forming melts, can be predicted, assuming that their glass phase is analogous to that of strong glasses. The disappearance of any Phase 2 at $T_{ts}$ in the melt true solution is determined. The Phase 2 formation temperature occurs at a branching temperature of thermal properties not far below $T_{ts}$.

## References

[1] P.S. Popel, V.E. Sidorov. Mater. Sci. Eng. A226-228 (1997) 237.

[2] P.S. Popel, V.V. Majekev, Zh. Phys. Chim. (USSR), 64 (1990) 568.

[3]. P.S. Popel, N.Y. Konstantinova, J. Phys. Conf. Series. 2008, Vol. 98, p. 062022.

[4] S. Wei, F. Yang, J. Bednarcik, I. Kaban, O. Shuleshova, A. Meyer, & R. Busch, Nature Comm. 4 (2013) 2083.

[5] C. Way, P. Wadhwa, and R. Busch, Acta Mater. 55 (2007) 2977.

[6] Q. Hu, H.C. Sheng, M.W. Fu, X.R. Zeng. J. Mater. Sci. 49 (2014) 6900.

[7] J.J.Z. Li, W.K. Rhim, C.P. Kim, K. Samwer, W.L. Johnson, Acta Mater. 59 (2011) 2166.

[8] S. Lan, M. Blodgett, K.F. Kelton, J.L. Ma, J. Fan, and X.L. Wang, Appl. Phys. Lett. 108 (2016) 211907.

[9] S. Wei, I. Gallino, R. Busch, and C.A. Angell, Nature Phys. 7 (2011) 178.

[10] H. Tanaka, R. Kurita, H. Mataki, Phys. Rev. Lett. 92 (2004) 025701.

[11] T.R. Kirkpatrick, D. Thirumalai, Phys. Rev. A. 31 (1985) 939.

[12] J. Souletie, J. Phys. France 51 (1990) 883.

[13] L. Berthier et al., Science 310 ( 2005) 1797.

[14] C.A. Angell, P.G. Wolynes, and V. Lubchenko [Eds]. Structural glasses and supercooled liquids. Wiley and Sons, Hoboken, NY, (2012) 237-278.

[15] M.I. Ojovan, K.P. Travis, R.J. Hand, J. Phys. Condens. Matter 19 (2007) 415107.

[16] M.I. Ojovan, J. Non-Cryst. Sol. 382 (2013) 79.

[17] J.F. Stanzione III, K.E. Strawhecker, R.P. Wool, J. Non-Cryst. Sol. 357 (2011) 311.

[18] R.P. Wool, J. Polymer. Sci. B 46 (2008) 2765.

[19] R.P. Wool, A. Campanella, J. Polymer Phys. B 47 (2009) 2578.

[20] D.S. Sanditov, J. Non-cryst. Sol. 385 (2014) 148.

[21] F. Albert et al., Science 352 (2016) 1308.

[22] R.F. Tournier, Physica B 454 (2014) 271.

[23] R.F. Tournier, Phys. B Condens. Matter 392 (2007) 79.

[24] R.F.Tournier, Chem. Phys. Lett. 651 (2016) 198.

[25] R.F. Tournier, J. Bossy, Chem. Phys. Lett. 658 (2016) 282.

[26] R.F. Tournier, Chem. Phys. Lett. 641 (2015) 9.

[27] D. Turnbull, Chem. Phys. 20 (1952) 411.

[28] R.F. Tournier, Revue de Metall. 109 (2012) 27.

[29] R.F. Tournier, Sci. Technol. Adv. Mater. 10 (2009) 014607.

[30] R.F. Tournier, Materials 4 (2011) 869.

[31] C-Y. Liu, J. He, R. Keunings, C. Bailly, Macromolecules 39 (2006) 8867.

[32] S. Mukherjee, Z. Zhou, J. Schroers, W.L. Johnson, and W.K. Rhim, Appl. Phys. Lett. 84 (2004) 5010.

[33] S.C. Glade, R. Busch, D.S. Lee, W.L. Johnson, R.K. Wunderlich, H.J. Fecht, J. Appl. Phys. 87 (2000) 7242.

[34] R. Busch, Y.J. Kim, W.L. Johnson, J. Appl. Phys. 77 (1995) 4039.

[35] K. Ohsaka et al., Appl. Phys. Lett. 70 (1997) 726.

[36] Z. Cernosek, J. Holubova, E. Cernoskova, M. Liska, Adv. Mat. 4 (2002) 489.

[37] A.R. Kurochkin, A.V. Borisenko, P.S. Popel, D.A. Yagodin, A.V. Okhapkin, High Temp. 51 (2013) 197.